\documentclass[lettersize,journal]{IEEEtran}
\usepackage{amsmath,amsfonts}
\usepackage{algorithmic}
\usepackage{algorithm}
\usepackage{array}
\usepackage{textcomp}
\usepackage{stfloats}
\usepackage{url}
\usepackage{verbatim}
\usepackage{graphicx}
\usepackage{cite}
\usepackage{subfigure}

\makeatother
\usepackage{xcolor}
\usepackage{multirow}
\usepackage{caption}
\usepackage{float}
\usepackage{soul}
\usepackage{graphicx}
\usepackage{textcomp}
\usepackage{wrapfig}
\usepackage{amsmath}
\usepackage{bm}
\usepackage{cite}
\usepackage{flushend}
\usepackage{subfigure}

\usepackage{graphicx}
\graphicspath{ {./images/} }

\def\s{\bm{s}}
\def\a{\bm{a}}
\def\t{_{t}}

\begin{document}

\title{DQLEL: Deep Q-Learning for Energy-Optimized LoS/NLoS UWB Node Selection}

\author{Zohreh~Hajiakhondi-Meybodi,~\IEEEmembership{Graduate Student Member,~IEEE,}
        Arash~Mohammadi,~\IEEEmembership{Senior Member,}
        Ming~Hou,~\IEEEmembership{Senior Member,}
        and~Konstantinos~N.~Plataniotis,~\IEEEmembership{Fellow,~IEEE}
\thanks{Z. HajiAkhondi-Meybodi is with Electrical and Computer Engineering (ECE), Concordia University, Montreal, Canada. (E-mail: z\_hajiak@encs.concordia.ca). A. Mohammadi (corresponding author) is with Concordia Institute of Information Systems Engineering (CIISE), Concordia University, Montreal, Canada. (P: +1 (514) 848-2424 ext. 2712 F: +1 (514) 848-3171, E-mail: arash.mohammadi@concordia.ca). M. Hou is with Defence Research and Development Canada (DRDC), Ottawa, Toronto, ON, M2K 3C9, Canada. (E-mail: ming.hou@drdc-rddc.gc.ca). K.~N.~Plataniotis is with Electrical and Computer Engineering (ECE), University of Toronto, Toronto, Canada. (E-mail: kostas@ece.utoronto.ca).}

\thanks{This Project was partially supported by Department of National Defence's Innovation for Defence Excellence \& Security (IDEaS)
program, Canada.}}



\maketitle

\begin{abstract}
Recent advancements in Internet of Things (IoTs) have brought about a surge of interest in indoor positioning for the purpose of providing reliable, accurate, and energy-efficient indoor navigation/localization systems. Ultra Wide Band (UWB) technology has been emerged as a potential candidate to satisfy the aforementioned requirements. Although UWB technology can enhance the accuracy of indoor positioning due to the use of a wide-frequency spectrum, there are key challenges ahead for its efficient implementation. On the one hand, achieving high precision in positioning relies on the identification/mitigation Non Line of Sight (NLoS) links, leading to a significant increase in the complexity of the localization framework. On the other hand, UWB beacons have a limited battery life, which is especially problematic in practical circumstances with certain beacons located in strategic positions. To address these challenges, we introduce an efficient node selection framework to enhance the location accuracy without using complex NLoS mitigation methods, while maintaining a balance between the remaining battery life of UWB beacons. Referred to as the Deep Q-Learning Energy-optimized LoS/NLoS (DQLEL) UWB node selection framework, the mobile user is autonomously trained to determine the optimal set of UWB beacons to be localized based on the 2-D Time Difference of Arrival (TDoA) framework. The effectiveness of the proposed DQLEL framework is evaluated in terms of the link condition, the deviation of the remaining battery life of UWB beacons, location error, and cumulative rewards. Based on the simulation results, the proposed DQLEL framework significantly outperformed its counterparts across the aforementioned aspects.
\end{abstract}


\section{Introduction}
\label{sec:introduction}
\IEEEPARstart{A}{s} an implication of recent  evolution of Internet of Things (IoT) applications~\cite{Li:2021}, reliable (trustworthy), accurate, and energy-efficient indoor navigation/localization systems are becoming increasingly important. In this context, various wireless technologies are available for indoor navigation/localization including but not limited to ZigBee~\cite{Loganathan:2019}, WiFi~\cite{Zhang:2020-1}, Radio-Frequency IDentification (RFID)~\cite{Motroni:2021}, Bluetooth Low Energy (BLE)~\cite{Hajiakhondi:2020,Hajiakhondi2020_1,Hajiakhondi2021}, and, Impulse Radio-Ultra Wide Band (IR-UWB)~\cite{Zafari:2019, Otim:2020, Wang:2020-1, Angarano:2020, Yang2021}. The latter, i.e., UWB is evolving as the wireless technology of the 2nd decade of the 21st century for providing reliable and accurate indoor localization with low latency~\cite{Raza:2017}. UWB systems use a high-bandwidth communication over a wide radio spectrum, which is equivalent to a narrow band signal in time domain. Therefore, they can resolve individual multi-path components. Consequently, the user's location can be estimated with high accuracy using the time information associated with the received UWB signal~\cite{Khalaf-Allah:2020}.

There are several approaches to extract location information from UWB signals among which Time Difference of Arrival (TDoA)~\cite{JHe2020} is an efficient solution. This is mainly due to the fact that the transmitter needs to send only one blink message to the receiver to be localized in the environment, which leads to a significant reduction of the UWB's power consumption~\cite{Wang:2019}. To localize and track mobile users based on the TDoA framework in a $2$-D area, time information from at least two UWB beacons~\cite{Zafari:2019} is required. 
TDoA-based localization approaches, however, are prone to NLoS error~\cite{Yassin:2016} due to presence of the obstacles within the indoor environment. In this context, Dai~\textit{et al.}~\cite{Dai:2013} analytically proved that using a small number of active beacons among all available ones not only improves the localization accuracy, but also mitigates the energy consumption of transmitting/receiving beacons.  Therefore, the main focus of recent researches~\cite{Wang:2020, Zhu:2016, Albaidhani:2019, Xie:2017,  Hadzic:2011, Hadzic:2014, Bel:2011, Courtay:2019, Albaidhani:2020, Wang:2016, Dai:2013, Zhang:2021 } has been shifted to design an efficient anchor node selection to increase the position accuracy by identifying beacons with LoS links instead of extracting location information from all beacons. Despite all the benefits that come from using existing anchor node selection frameworks, there are still critical challenges ahead. On the one hand, energy efficiency is compromised in scenarios where all beacons require to transmit/receive signals for LoS/NLoS identification. On the other hand, extracting  location  information  from  just  beacons  with  LoS condition results in quickly draining the battery of certain LoS beacons. Furthermore, analytical anchor node selection frameworks developed based on fixed mathematical models fail to cope with the  dynamic nature of indoor environments, such as unknown/varying adverse environmental conditions. Capitalizing on the aforementioned discussion, the paper focuses on the issues of the location error caused by NLoS connections and the unbalanced energy consumption of UWB beacons. We propose an innovative Deep Q-Learning-based Energy-optimized LoS/NLoS (DQLEL) UWB node selection framework, that efficiently cope with the dynamic nature of indoor environments. In what follows, to understand the state-of-the-art in this area and seek potential solutions, we first review the relevant literature.



\noindent
\textbf{Related Work}:
Anchor node selection in indoor localization is used to improve the network's performance by defining a set of criteria for selecting a subset of beacons with the highest utilities. Therefore, a set of UWB beacons with low location error, i.e., LoS connections, will be selected for monitoring/tracking users' locations instead of extracting location information from all available beacons. With the application to the Wireless Sensor networks (WSNs), Zhu~\textit{et al.}~\cite{Zhu:2016}
introduced the Link Condition Indicator (LCI) as a metric for anchor node selection in wireless sensor networks to mitigate the destructive effect of the NLoS conditions. Dai~\textit{et al.}~\cite{Dai:2013} proposed a near optimal solution for anchor node selection in Wireless Network Localization (WNL). Zhang~\textit{et al.}~\cite{Zhang:2021} mitigated the location error caused by NLoS connections in an indoor environment using topological unit, which is defined as the topological relationship between nodes.

Within the UWB-based indoor localization context, Wang~\textit{et al.}~\cite{Wang:2016} introduced multiple non-overlapping Effective Localization Areas (ELAs), where each ELA consists of four UWB beacons. Accordingly, after evaluation of the  Channel Impulse Response (CIR) from all beacons, two out of four beacons were involved in the localization if the established links are LoS connections.  Albaidhani~\textit{et al.}~\cite{Albaidhani:2019} proposed an UWB node selection based on the Mean Squared Error (MSE) metric, where UWB beacons are clustered into different groups. Then, the group with the lowest MSE is selected to be used for localization by using the Weighted Least Square (WLS) method. Albaidhani~\textit{et al.}~\cite{Albaidhani:2020} introduced another evaluation metric for selecting the best set of UWB beacons, named Geometric Dilution of Precision (GDOP), and showed the superiority of the GDOP in comparison to the MSE metric in terms of the location accuracy. Despite the fact that all the existing works related to this area alleviate the node's energy consumption and location error, there is still a challenge ahead, which is keeping a trade-off between the remaining battery life of beacons and localization accuracy. Extracting location information from just beacons with LoS condition, however, improves the location accuracy, it results in quickly draining the battery of certain LoS beacons. Then, it is required to replacement them with the new beacons or using other beacons with NLoS conditions. Due to the limited battery life of beacons, therefore, it is of paramount importance to keep a balance between the remaining battery life of all beacons. \textit{This further motivates us to develop an energy-optimized LoS/NLoS UWB node selection solution to boost the location accuracy while maintaining a balance between the battery life of UWB beacons.}

\noindent
\textbf{Significance}:
We aim to address the aforementioned unbalanced energy consumption issue, the location error caused by the NLoS connections, and the time varying behavior of indoor environments. In this context, we investigate an autonomous and homogeneous indoor localization framework including fixed and known location UWB beacons covering the area of interest. The goal is to monitor/track user's movement with high accuracy in the presence of NLoS condition. Due to the limited battery life of UWB beacons and the computational complexity of the localization phase, it is essential to minimize the number of cooperating nodes for localization. The novel approach we are taking here is to train the mobile user to be localized through the optimal UWB beacons with LoS links, while maintaining a balance between the remaining battery life of all beacons. In summary, the paper makes the following key contributions:
\begin{itemize}
\item Due to the reflective obstacles such as walls and human body, indoor environments suffer from NLoS connections, which degrade the location accuracy exponentially. Without applying complex NLoS mitigation methods, we introduce an autonomous localization framework, where the mobile user is trained to be localized by UWB beacons with LoS conditions at each time/location.
\item In such a scenario that a set of UWB beacons with LoS condition mostly contribute to localization, their batteries would be fully drained in a short period of time. It is, therefore, imperative to consider the remaining battery life of beacons as a selection criteria. 
Analytical anchor node selection frameworks are unable  to  cope  with  the  dynamic  nature  and  time-varying  behavior of indoor environments. We, therefore, target development of an adaptive anchor selection framework that efficiently cope with the dynamic nature of indoor environments. Despite the surging interest in the anchor node selection frameworks, there is no Reinforcement Learning (RL)-based framework concerning a trade-off between the remaining battery life of UWBs and localization accuracy. The proposed DQLEL framework addresses this gap via a RL-based formulization with the goal of maintaining a balance between the location error by selecting UWB beacons with the LoS condition and the remaining battery life of UWB beacons.
\end{itemize}
The effectiveness of the proposed DQLEL framework is evaluated through comprehensive simulation studies in terms of the location error, the mean deviation of UWB's remaining  battery life, the link condition, and the cumulative rewards. Simulation results illustrate the efficiency of the proposed DQLEL scheme in comparison to its state-of-the-art counterparts over all the aforementioned aspects.

 
The rest of this paper is organized as follows: In Section~\ref{Sec:2}, the system model and the problem description are provided. Section~\ref{Sec:3} deals with introducing the proposed  DQLEL UWB node selection framework. Simulation results are presented in Section~\ref{Sec:4}. Finally, in Section~\ref{Sec:5}, an overview of the results and concluding remarks are presented.
\begin{table*}[ht]
\centering
\caption{List of Notations.}\label{table1}
\begin{tabular}{  |c |l|| c|l|} \hline
\textbf{Notation} & \textbf{Description} & \textbf{Notation} & \textbf{Description}\\ \hline
$N$ &  Total number of UWB beacons & $P_{COR}$ & Correlator branch  power consumption \\
$N_u$ &  Number of available beacons & $P_{ADC}$ & ADC power consumption\\
$R_i$ &  Reception range of $\text{UWB}_i$ & $P_{LNA}$ & LNA power consumption\\
$B_{i,t}$ &  Battery life of $\text{UWB}_i$ at time slot $t$& $P_{VGA}$&  VGA power consumption\\
$\tau_{i}$ & First peak of the estimated CIR of $\text{UWB}_i$ &$P_{GEN}$& Pulse generator power consumption\\
$(x_t,y_t)$ & User's location at time slot $t$ & $P_{SYN}$ & Synchronization power consumption\\
$(x_i,y_i)$ & Location of $\text{UWB}_i$ at time slot $t$& $P_{EST}$ &  Channel estimator power consumption \\
$P_r$ & Power consumption of an UWB receiver &$E_L$ &Payload  energy consumption\\
$\rho_t$ & Pulse coefficient  &$E_O$&  SP and PHR energy consumption\\
$\rho_c$ & Symbol repetition parameter & $E_{RX}$ & Energy consumption for receiving one packet\\
$\rho_r$ & Demodulation parameter& $E$& Total energy consumption in reception session\\
$L_{SP}$ & Number of SP symbols &$E_{IPS}$& Energy consumption during $T_{IPS}$ \\
$L_L$ & Number of bits in the payload & $E_{ACK}$ &Acknowledgement power consumption\\
$L_{PHR}$ & Number of bits in the PHR &$T _{tr}$& Transient session\\
$R_{base}$& Fixed base data rate & $T_{on}$& Time duration  for receiving one packet \\
$R_c$ & Coding rate& $T_{onL}$&  Time duration of the payload \\
$N_p $ & Coding parameter &$T_{PHR}$ & Time duration of the PHR \\
$T_{ACK}$& Acknowledgement time &$T_{SP}$ & Time duration of the SP \\
$T_{IPS}$ & Inter packet space time duration & $M$ & Number of RAKE fingers\\ \hline
\end{tabular}
\vspace{-.2in}
\end{table*}

\section{System Model and Problem Description} \label{Sec:2}
Our research scenario is a corridor inside a building (e.g., an office  or a hotel building), consisting of $N$ rooms where each room is equipped with a synchronized UWB beacon as the receiver node, denoted by $UWB_i$, $i=1, \ldots ,N$. We also consider a mobile user as the transmitter node, who randomly moves within the corridor. 
The reception range (assumed to be equal) and the battery life of UWB beacons at time slot $t$ are denoted by $R_i$ and $B_{i,t}$, for ($1 \leq i \leq N$), respectively. Due to the existence of reflective obstacles such as walls and movement of other users, such indoor environments suffer from multipath, shadowing, and pathloss effects, which are known by NLoS links. Therefore, the transmission link between UWB beacons and the user at each location can be either LoS or NLoS. Despite the ToA approach, where a strict synchronization is required between the transmitter and the receiver, only synchronization between UWB beacons is required in the TDoA. By relaxing the assumption that all UWB beacons are synchronized, our objective is to train the user to be localized through an optimal set of UWB beacons with LoS links, without draining the battery of certain UWB beacons. In this section, we  present the wireless signal model of the IR-UWB standard and formulate the transmitted signal, wireless channel, and the received signal to extract the user's location through the time information. Afterwards,  the energy consumption model of the UWB beacons as the receiver nodes will be introduced. A summary of the notations used hereinafter is provided in~Table~\ref{table1}.

\vspace{.025in}
\noindent
\textit{A. IR-UWB Wireless Signal Model}

In the IR-UWB technology, sequences of short time-domain impulses transmit over a high-bandwidth radio spectrum  resulting in an improvement in data rate and localization accuracy for short-range communication. In this paper, the original bit stream, denoted by $b_n$, is modulated based on the Pulse Amplitude Modulation (PAM) method~\cite{Gunturi:2017}, which is known as one of the efficient IR-UWB modulation schemes. In this case, the baseband version of the transmitted UWB signal $s_u(t)$ is expressed~as
\begin{equation}\label{e1}
s_u(t)= p_n \psi(t), ~~~\text{where} ~~ p_n=
\left\{ \begin{array}{ll}
1 & b_n=1\\
 -1  & b_n=0,\\
\end{array} \right.
\end{equation}
where $\psi(t)$ denotes the $5^{\text{th}}$ derivative of the Gaussian signal and $p_n$ is the polarity factor. In the indoor environments, UWB beacons, typically, receive a number of phase delayed and power attenuated versions of the transmitted UWB signal $s(t)$, affected by Additive White Gaussian Noise (AWGN). The received signal by $UWB_i$, denoted by $r_i(t)$, therefore, can be expressed as
\begin{equation}\label{e2}
r_i(t) =\sum_{k=1}^{N(t)} \rho_k(t,\tau) s(t-\tau_k(t)-\tau_{i})+n(t),
\end{equation}
where $N(t)$ represents the number of detachable paths. Terms $\rho_k(t,\tau)= \beta_k(t,\tau) \exp{(j\Phi_k(t,\tau))}$ and $\tau_k(t)$ denote the attenuation and the delay of the $k^{\text{th}}$ path, where $\beta_k(t,\tau)$ and $\Phi_k(t,\tau) $ represent the path amplitude and phase, respectively. Term $\tau_{i}$ represents the ideal time information, which is equal to $\tau_{i}=d_i/c$, where $d_i$ is the distance between the user and $UWB_i$, and $c$ is the speed of light, about $3 \times 10^8$ $m/s$. Finally, term $n(t)$ represents the AWGN channel, which is modelled by $n(t) \sim \mathcal{N}(0,\sigma ^2)$. Note that, $\beta_k(t,\tau)$ is the Nakagami-m random variable as the small-scale channel coefficient of the link between the user and the UWB beacon, where $m$ indicates the degree of fading severity. By assuming that there is a strong LoS path between the transmitter and the receiver, the radio propagation is modelled by Rician fading channel model, otherwise, it would be Rayleigh fading.  More specifically, $m=1$ represents the Rayleigh fading, while $m > 1$ is for Rician channel model. Under the Nakagami-m fading assumption, the small-scale channel gain denoting by $|\zeta_k|^2 \sim \Gamma(m,1/m)$ is a normalized independent and identically distributed (i.i.d.) Gamma random variable. The Probability Density Function (PDF) of the power fading is expressed as
\begin{equation}\label{e3}
f_{|\zeta_k|^2 }(x)=\dfrac{m^m}{\Gamma(m)}x^{m-1}e^{-mx},~~~ x>0
\end{equation}
where $\Gamma(\cdot)$ is the Gamma function. In an UWB positioning system, the user's location can be obtained from the estimated CIR, which is expressed as
\begin{equation}\label{e4}
h(t,\tau)=\sum_{k=1}^{N(t)}\rho_k(t,\tau) \delta(t-\tau_k(t)-\tau_{i}),
\end{equation}
where the first maximum peak of the estimated CIR in the LoS condition is associated with the delay of the first path. TDoA scheme determines the location of users by calculating the time difference between the transmitted signal from the user and the signals received by at least two UWB beacons in a 2-D indoor environment. In such a scenario that the corresponding user is located in the reception range of $UWB_i$ and $UWB_j$, the TDoA information, denoted by $T_{ij}$, is given by
\begin{eqnarray}\label{e5}
\lefteqn{~~~~~T_{ij} =\tau_{i}-\tau_{j}= \nonumber}\\
&&\!\!\!\!\!\!\!\!\!\!\!\!\!\!\!\!\!\!\!\!\frac{\sqrt{\big(x_t-x_i\big)^2+\big(y_t-y_i\big)^2}-\sqrt{\big(x_t-x_j\big)^2+\big(y_t-y_j\big)^2}}{c},
\end{eqnarray}
where $\tau_{i}$ and $\tau_{j}$ represent the first peak of the estimated CIR of  $UWB_i$ and $UWB_j$. In addition, ($x_t, y_t$), ($x_i, y_i$), and ($x_j, y_j$) denote the locations of the corresponding user at time slot $t$, $UWB_i$, and $UWB_j$, respectively. This completes the description of the UWB wireless signal model. Next, we introduce the UWB power consumption model.

\vspace{.1in}
\noindent
\textit{B. IR-UWB Energy Consumption Model}

Due to the limited battery life of UWB beacons as the receiver nodes, it is crucial  to extend the lifetime of beacons. Toward this goal, first we present the energy consumption of UWB beacons~\cite{Wang:2010}. Then, we propose a DRL framework to involve beacons in the localization in such a fairness scenario that the battery life of almost all beacons is the same at each time slot. In this regard, the power consumption of an IR-UWB receiver $P_r$ is expressed as
\begin{equation}\label{e6}
P_r = P_d +P_{n},
\end{equation}
where $P_d$ and $P_{n}$ represent the circuit components' power consumption associated with the detection scheme, and the rest of the components, respectively. In such a case, $P_d$ and $P_{n}$ are calculated as follows
\begin{eqnarray}
P_d &=&M P_{COR}+\rho_c P_{ADC}+ P_{LNA}+ P_{VGA},\label{e7}\\
\text{and }\quad P_{n}&=&\rho_r (P_{GEN}+P_{SYN}+P_{EST}) \label{e8}.
\end{eqnarray}
where $P_{COR}$, $P_{ADC}$, $P_{LNA}$, and $P_{VGA}$ denote the power consumption of the correlator branch including a mixer and an integrator, the Analog-to-Digital Converter (ADC), the Low Noise Amplifier (LNA), and the Variable Gain Amplifier (VGA), respectively. Term $M$ denotes the number of  RAKE fingers at the receiver side, which is assumed to be $1$. Symbol repetition scheme, including Hard Decision (HD) combining and Soft Decision (SD) combining is denoted by $\rho_c$, where $\rho_c=1$ is used for SD and $\rho_c=0$ is for HD combining. In addition,  $P_{GEN}$, $P_{SYN}$, and $P_{EST}$ represent the power consumption associated with the pulse generator, the synchronizer, and the channel estimator, respectively. Finally, $\rho_r$ is an structural parameter, where $\rho_r=1$ is related to the coherent demodulation while $\rho_r=0$ is for noncoherent demodulation.

\begin{figure}[t!]
\centerline{\includegraphics [scale = 0.55] {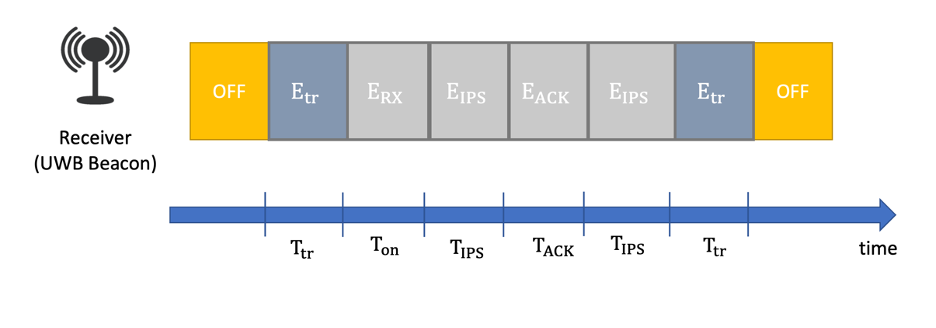}}
\caption{\footnotesize Energy consumption of UWB beacons during one packet reception.}
\label{Fig4}
\end{figure}

The IR-UWB data packet is constructed by: (i) Synchronization Preamble (SP); (ii) PHY-Header (PHR), and; (iii) Payload components. Therefore, the energy consumption of receiving one data packet consists of the energy consumed on the payload, denoted by $E_L$, and the energy consumption associated with delivering the SP and PHR, denoted by $E_O$, where $E_O$ is expressed as
\begin{equation}\label{e2}
E_O= P_r \big(T_{SP} + T_{PHR}\big) =  P_r\underbrace{\frac{(L_{SP}+\dfrac{L_{PHR}}{R_c})}{R_{base}}}_{T_O},
\end{equation}
with the assumption that there are $L_{SP}$ symbols in the SP, and $L_{PHR}$ bits in the PHR, respectively. The time duration of the PHR, and the SP are represented by $T_{PHR}$, and $T_{SP}$, respectively. Term $R_c = 1/N_p$ is the coding rate, where $N_p$ is the coding parameter, which must be an odd number. Finally, $R_{base}$ is the fixed base data rate. The energy consumption for receiving the payload, containing $L_L$ information bits, denoted by $E_L$, is expressed~as
\begin{eqnarray}\label{e2}
E_L &=& \rho_t(M P_{COR}+\rho_c P_{ADC}+ P_{LNA}+ P_{VGA})T_{onL} \nonumber \\
&+& \rho_r (P_{GEN}+P_{SYN})T_{onL},
\end{eqnarray}
where $\rho_t$ represents the pulse coefficient, and $T_{onL}$ denotes the time duration of the payload. The energy consumption to receive one packet within time $T_{on}$ is $E_{RX}=E_O+E_L$, which can be expressed as
\begin{equation}\label{e2}
T_{on}=T_{SP} + T_{PHR}+T_{onL}=\frac{(L_{SP}+\dfrac{L_{PHR}}{R_c})}{R_{base}}+\dfrac{L_L}{R_b R_c},
\end{equation}
%
where $R_b$ denotes the bit rate. Since our focus is to manage the energy consumption of UWB beacons, the amount of energy consumed by users as the transmitter is not modeled. As it can be seen from Fig.~\ref{Fig4}, the total energy consumed by UWB beacons during each packet reception session can be expressed~as
\begin{equation}\label{e12}
E=2E_{tr}+E_{RX}+2E_{IPS}+E_{ACK},
\end{equation}
where $E_{IPS}=\rho_rP_{SYN}T_{IPS}$ and $E_{tr}=\rho_r P_{SYN} T_{tr}$. In this case, $T_{tr}$ represents the time that the UWB beacon switches from the sleep state to an active state for receiving a data packet, and $T_{IPS}$ denotes the inter packet space time duration. Moreover, $E_{ACK}$ is calculated as
\begin{equation}\label{e2}
E_{ACK}=(L_{SP}+\dfrac{L_{PHR}}{R_c})E_p+P_{SYN}T_{ACK},
\end{equation}
where $T_{ACK}=T_O$ is the time duration when the UWB beacon listens for an ACK acknowledgement from the corresponding user. This completes the description of the UWB power consumption model. Next, we introduce our proposed DQLEL UWB node selection framework.

\setlength{\textfloatsep}{0pt}

\section{DQLEL UWB Node Selection} \label{Sec:3}
In this section, we first briefly introduce the required background on RL, and then present the proposed DQLEL.

\vspace{.1in}
\noindent
\textit{A. RL Background}

RL model is a type of Machine Learning (ML) technique, where an agent interacts with its  environment to learn the optimum action in a given state from a set of given actions. After each interaction, the agent receives a feedback from the environment, that could be a reward or a punishment to update its state accordingly. This whole process is known as the Markov Decision Process (MDP), which includes a set of $\mathcal{A}$ actions, a set of states $\mathcal{S}$, a transition function $\mathcal{T}$, and a reward function, denoted by $\mathcal{R}$. According to the transition function $\mathcal{T}(s_t, a_t, s_{t+1})$ and the reward function $r_t=\mathcal{R}(s_t, a_t)$, an action $a_t \in \mathcal{A}$ at time slot $t$ in any state $s_t \in \mathcal{S}$ results in a new state $s_{t+1} \in \mathcal{S}$ at time slot $t+1$. The optimum policy, denoted by $\pi^*$, is a policy that leads to achieving the maximum accumulated rewards during interactions, expressed~as
\begin{eqnarray}
\pi^* = \arg \max \limits_{\substack{\pi}} \mathbb{E}_{\pi}  \Big \{ \sum \limits_{t=0}^{H-1} \gamma^{t} r_{t+1} | \bm{s}_0 = s \Big \},
\end{eqnarray}
where $\gamma \in [0,1]$ and $H$ represent the discount factor and the number of finite episodes in MDP. While the low value of $\gamma$ maximizes the short-term rewards, a higher one leads to increasing the long-term rewards. One of the most widely used value-based and model-free RL algorithms is the Q-learning framework, where the Q-value of the action $a_t$ and the state $s_t$ of the agent at time slot $t$, denoted by $Q(s_t, a_t)$, is calculated~as
\begin{equation}
Q(s_t, a_t) = \mathbb{E}_{\pi} \Big \{ \sum \limits_{t=0}^{H-1} \gamma^{t} r_{t+1} | s_0 = s, a_0= a, a_t = \pi({s_t}) \Big\}.\label{nEq:33}
\end{equation}
Then, the updated Q-value in each time slot is obtained as
\begin{equation}
Q(s_t, a_t) \leftarrow  (1-\lambda) Q(s_t, a_t)+ \lambda (r_t+ \gamma \max Q(s_{t+1}, a_{t+1})),
\end{equation}
where $\lambda \in [0,1]$ is the learning rate. In such a scenario that the action and the state spaces are finite, the Q-learning approach performs efficiently by constructing a Q-table to look up and update the Q-value associated with an action-state pair. By considering the fact that the number of actions and state-space are infinite in the proposed DQLEL framework, it is essential to apply Deep Q-Learning (DQL) approaches as an approximator. Therefore, we use the Convolutional Neural Network (CNN) as one of the widely used Deep Learning (DL) methods to estimate Q-values as described below.

\begin{figure}[t!]
\centerline{\includegraphics [scale = 0.5] {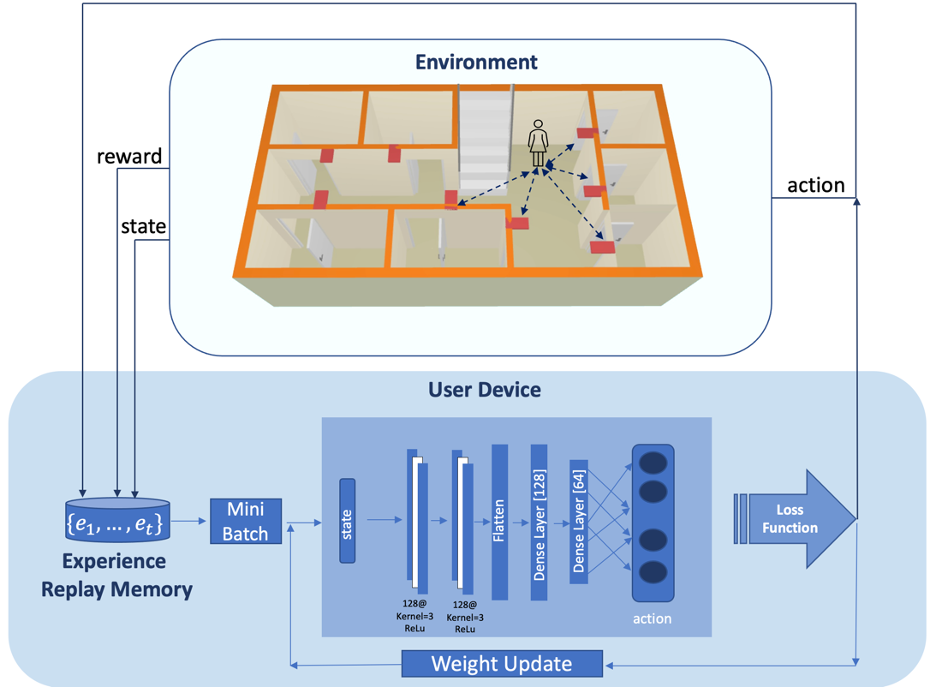}}
\caption{\footnotesize The block diagram of the proposed DQLEL  framework.}\vspace{.1in}
\label{Fig3}
\end{figure}

\vspace{.1in}
\noindent
\textit{B. DQLEL Framework}

As stated previously, extracting time information from NLoS links leads to an increase in the location error. At the same time,  extracting time information from all beacons is inefficient  from energy consumption perspective. The proposed DQLEL framework capitalizes on these facts aiming to autonomously train  the mobile user to find energy-optimized UWB beacons with LoS links at each location. Considering the fact that applying a complex RL model is energy consuming, it is commonly assumed that UWB beacons transmit their sensory data to the central server~\cite{Lei:2020} to perform the DQLEL framework. As it can be seen from Fig.~\ref{Fig3}, the DQLEL framework consists of the following main components:

\vspace{.1in}
\noindent
\textbf{(i) Agent:} The mobile user acts as the agent within the DQLEL framework and interacts with the environment based on a set of given actions defined below.

\vspace{.1in}
\noindent
\textbf{(ii) Action-Space:} The action space in the DQLEL  framework refers to the selection of a set of UWB beacons to determine the location of the mobile user. Since two UWB beacons are required to determine the user's location, the selected action is a vector, denoted by $\mathbf{a}=[a_i,a_j]$, where $a_i$ and $a_j$ represent $UWB_i$ and $UWB_j$, respectively. Therefore, the cardinality of the action space $N_a$ is defined as follows
\begin{equation}
N_a= \frac{N_u!}{(N_u-2)! \enspace 2!},
\end{equation}
where $N_u$ denotes the number of UWB beacons in the vicinity of the user. By taking an action, the estimated location of the user is calculated. Note that the candidate UWBs' link condition, i.e., LoS or NLoS feature has a great impact on the location accuracy. Therefore, the location error at time slot $t$, denoted by $\mathcal{ER}_t$, is calculated as
\begin{equation}\label{e18}
\mathcal{ER}_t=\sqrt{(x_t-x_{es,t}^{(i,j)})^2+(y_t-y_{es,t}^{(i,j)})^2},
\end{equation}
where $(x_t,y_t)$ is the exact user's location at time slot $t$ and $(x_{es,t}^{(i,j)},y_{es,t}^{(i,j)})$ denotes the estimated user's location, which is obtained by $UWB_i$ and $UWB_j$. Note that selecting UWB beacons with LoS links leads to a remarkable reduction in the value of $\mathcal{ER}_t$.

\vspace{.1in}
\noindent
\textbf{(iii) State-Space:} The action is selected based on the current state of the system $\mathbf{s}_t$ at time slot $t$. Each state consists of the user's location $(x_t,y_t)$, and the UWB beacons' battery life $\mathbf{B}_t$. More specifically, $\mathbf{s}_t$ consists of the following components:
\begin{itemize}
\item $(x_t,y_t)$: Location of the user. Following Reference~\cite{Li:2020}, the indoor environment is discretized into $N_l= N_x \times N_y$ points, where $x_t$ and $y_t$ at time slot $t$ are obtained as
\begin{eqnarray}\label{e19}
x_t&=& x_{t-1} + x_m\label{e7},~~~~~0 \leq x_t \leq N_x\\
\text{and } y_t&=& y_{t-1} + y_m,~~~~~0 \leq y_t \leq N_y \label{e20}.
\end{eqnarray}
where $(x_{t-1},y_{t-1})$ denotes the user's location at time slot $t-1$, and $ x_m, y_m \in \{-1,0,1\}$ indicate the user's movement~\cite{Li:2020} and are selected randomly.

\item $\mathbf{B}_t$: Vector $\mathbf{B}_t = [B_{1,t}, \ldots, B_{N_u,t}]$ illustrates the battery life of $UWB_i$, for ($1 \leq i \leq N_u$). By taking the action $\a$, the battery life of two involving beacons are updated as
\begin{equation}
B_{i,t}=B_{i,t-1}-E,
\end{equation}
where $E$ denotes the energy consumption of UWB beacon in the reception session, obtained by Eq.~\eqref{e12}.
\end{itemize}

\vspace{.1in}
\noindent
\textbf{(iv) Reward:} The reward function in the proposed DQLEL framework is defined in such a way that the user selects UWB beacons with LoS links, while maintaining the load balance between UWB beacons. To provide a fairness UWB connection scheduling, the mean deviation of UWB energy consumption is introduced as the load balance metric. In this regard, the deviation of remaining battery of UWB beacons at time slot $t$, denoted by $\mathcal{MD}_{t}$, is the distance between the battery of all beacons at time slot $t$ and the average one, given by
\begin{equation}\label{e22}
\mathcal{MD}_{t}= \sqrt{\dfrac{1}{N_u-1}\sum \limits_{i=1}^{N_u} \dfrac{(B_{i,t}-\overline{B}_{t} )^2}{\overline{B}_{t} ^2}} ,
\end{equation}
where $\overline{B}_{t}$ represents the average UWB battery life at time slot $t$. 
The reward function $\mathcal{R}(\s\t, \a\t)$ associated with the link condition and UWB energy consumption at time $t$ is defined~as

\begin{eqnarray} \label{e23}
&&\!\!\!\!\!\!\mathcal{R}(\s\t, \a\t) = \\&&\left\{\begin{array}{ll}
 10,~~~\mbox{\textbf{C1.}~ $\mathcal{MD}_{t} \leq \mathcal{MD}_{th}$, $c_{i,t}=1$,$c_{j,t}=1$,}\\
5,~~~~\mbox{\textbf{C2.}~~ $\mathcal{MD}_{t} > \mathcal{MD}_{th}$, $c_{i,t}=1$,$c_{j,t}=1$,}\\
-5,~~\mbox{\textbf{C3.}~ $\mathcal{MD}_{t} \leq \mathcal{MD}_{th}$, $c_{i,t}=0$,$c_{j,t}\in \{0,1\}$,}\\
-10,~\mbox{\textbf{C4.}~ $\mathcal{MD}_{t} > \mathcal{MD}_{th}$, $c_{i,t}=0$,$c_{j,t}\in \{0,1\}$.}\end{array}\right. ,\nonumber
\end{eqnarray}
where $c_{i,t}$ and $c_{j,t}$ indicate the links' condition associated with the given action $\a$. In DQLEL, we define a ($N_u \times N_l$) link condition matrix, denoted by $\textbf{C}$, where the $t^{\text{th}}$ column illustrates the link condition established between the user at point $(x_t,y_t)$ and all UWB beacons. Then, $c_{i,t}=1$ if the link between the user at location $(x_t,y_t)$ and the $i^{\text{th}}$ UWB beacon is LoS, otherwise $c_{i,t}=0$. Finally, $\mathcal{MD}_{th}=\varepsilon E$ is a small pre-defined value, where $0<\varepsilon<1$. A lower $\varepsilon$ value reduces the gap between the remaining battery life of beacons, extending the network's life time. Term $\mathcal{MD}_{t} \leq \mathcal{MD}_{th}$ provides a load balance between all UWB beacons to prevent any beacons' batteries from being completely drained. According to Eq.~\eqref{e23}, both location accuracy and the balanced energy consumption of beacons are considered as the reward function, where the former is stated in terms of the link condition $c_{i,t}$ and $c_{j,t}$ and the latter one is expressed as $\mathcal{MD}_{t}$. More precisely, the reward function in Eq.~\eqref{e23} illustrates the following four connection types for the selected pair of UWB beacons: (\textbf{C1.}) Energy-optimized with LoS links; (\textbf{C2.}) Non energy-optimized with LoS links; (\textbf{C3.}) Energy-optimized with NLoS links, and; (\textbf{C4.}) Non energy-optimized with NLoS links. An energy-optimized link is referred to the action where the battery life deviation of all UWBs is less than a pre-determined threshold $\mathcal{MD}_{th}$; otherwise, it is called a non energy-optimized connection. Moreover, $c_{i,t}=0$, $c_{j,t}\in \{0,1\}$ means that at least one link of the selected pair of UWB beacons is NLoS, leading to a remarkable location error. Therefore, to achieve a high location accuracy, both UWB beacons should be LoS, i.e., $c_{i,t}=1$, $c_{j,t}=1$. Note that, in the worst-case scenario, the selected pair are non energy-optimized with NLoS links. Our goal is to increase the energy-optimized with LoS connection type, while reducing the non energy-optimized with NLoS connections. After selecting two UWB beacons with the largest reward function $\mathcal{R}(\s\t, \a\t)$, the connection information associated with the corresponding action and state are stored in the memory replay of the proposed DQLEL model. Due to the infinite state-action space, we use the CNN architecture as a non-linear approximator in the Q-learning model to approximate the Q-value of each state-action pair.

The CNN module of the DQLEL framework consists of two 1-dimensional convolutional layers, consisting of $128$ filters, each with the size of $3$ and with the ReLU activation function. There are also two Fully Connected (FC) layers, where the first one consists of $128$ ReLU units and the latter has $64$ ReLU units. The number of input is equal to the size of the state-space, which is equal to $(4 + N_u)$, and the number of output layer is equal to the size of the action space. Moreover, the activation function of the output layer is softmax. To maintain a trade-off between the exploration and the exploitation of the DQLEL framework, a variable $\epsilon$ is assumed for the $\epsilon$‐greedy action selection policy. The maximum value of $\epsilon$, denoted by $\epsilon_{max}$, is equal to $1$, gradually decreasing with time by $\Delta \epsilon=\dfrac{\epsilon_{max}-\epsilon_{min}}{N_{epoch}} $ until a steady state is reached, where $\epsilon_{min}=0.01$, and $N_{epoch}$ is the total number of epochs, equal to $500$ in this work. In such a scenario, the random action $\a_t$ is selected at time slot $t$ with the probability of $\epsilon$.

A replay memory is used to retrain the CNN model for previously observed state-action pairs and their corresponding rewards. Therefore, $\beta$ number of state-action pairs at time slot $t$, denoted by $\boldsymbol{\phi}_t=[\boldsymbol{s}_{t-\beta},\boldsymbol{a}_{t-\beta},\ldots,\boldsymbol{a}_{t-1},\boldsymbol{s}_t]$, are used as the input of the CNN to estimate $Q(\boldsymbol{\phi}_t, \boldsymbol{a}_t|\xi_t)$, where $\xi_t$ denotes the filter weight at time slot $t$. The experience memory pool is denoted by $D=\{\boldsymbol{e}_1,\ldots,\boldsymbol{e}_t \}$, where $\boldsymbol{e}_t=(\boldsymbol{\phi}_t,\boldsymbol{a}_t,r_t,\boldsymbol{\phi}_{t+1})$. To update the weight parameter $\xi_t$ using the Stochastic Gradient Descent (SGD) method, the state sequence in replay buffer $\textbf{e}_m$ is selected at random. Given the value of $\xi_t$, the goal is to obtain the optimal action in each time slot, which is obtained by minimizing the following loss function
\begin{equation}\label{e24}
\mathbb{L}(\xi_t)=\mathbb{E}_{{\boldsymbol{\phi}}_t},\boldsymbol{a}_t,r_t,\boldsymbol{\phi}_{t+1} \left [ \big(Q_{T}-Q(\boldsymbol{\phi}_t,\boldsymbol{a}_t| \xi_{t+1}) \big)^2 \right] ,
\end{equation}
where $Q_{T}$ is the target optimal Q-function, expressed as
\begin{equation}
Q_{T}=r_t+\gamma \max \limits_{\substack{\a\t^{'}}} Q(\boldsymbol{\phi}_{t+1},\boldsymbol{a}_{t}^{'}|\xi_{t-1}).\label{nEq:39}
\end{equation}
According to the $\epsilon$-greedy algorithm, the best action $\boldsymbol{a}_t^{*}$ for the state $\boldsymbol{s}_t$ is chosen from the set of Q-functions with the probability of ($1-\epsilon$) as follows
\begin{equation}
\boldsymbol{a}_t^{*}= \arg \max \limits_{\substack{\a\t^{'}}} Q(\boldsymbol{\phi}_t, \boldsymbol{a}_{t}^{'}).
\end{equation}
Given the action $\boldsymbol{a}_t^{*}$, two UWB beacons associated with action $\boldsymbol{a}_t^{*}$  are involved to track the user's location at time slot $t$. Eventually, the new experience $\{\boldsymbol{\phi}_t,\boldsymbol{a}_t,r_t,\boldsymbol{\phi}_{t-1} \}$ is stored in the replay memory by the agent.

\vspace{.05in}
\noindent
\textit{C. Computational Complexity}

In this Subsection, we compute the computational complexity of the CNN as the learning method of the proposed DQLEL framework. Generally speaking, the computational complexity of a CNN model with $\mathcal{N}_l$ number of convolutional layers, where each layer includes $F_l$ filters with size $W_l^f \times L_l^f$, is
\begin{equation}\label{e41}
\mathcal{N}=\sum \limits_{l=1}^{\mathcal{N}_l} F_{l-1} W_l^f L_l^f F_l  W_l^o  L_l^o,
\end{equation}
where $F_{l-1}$ and $F_l$ represent the number of input channels and filters corresponding to the $l^{\text{th}}$ layer, respectively. In addition, $W_l^o$ and $L_l^o$ denote the width and the length of the output, calculated as follows
\begin{eqnarray} W_l^o&=&\dfrac{W_{l-1}^o- W_l^f+2P_l}{S_l}+1,\\\label{e42}
\quad \text{and}~L_l^o&=&\dfrac{L_{l-1}^o- L_l^f+2P_l}{S_l}+1,\label{e43}
\end{eqnarray}
where $S_l$ and $P_l$ represent the size of stride and padding layers of the $l^{\text{th}}$ layer, respectively. Moreover, there are $\mathcal{N}_{fc}$ number of fully connected layers  for estimating the Q-value associated with each action. Considering that the pooling and fully connected layers only take up $5-10\%$ of the computational time~\cite{Aref:2019}, their impact on the computational complexity of the CNN can be negligible. Accordingly, the value of $F_0 \times W_0^{o} \times L_0^{o}$ is equal to $\beta l_s$ in the DQLEL framework, where $\beta$ is the temporal memory depth, and $l_s$ represents the length of the state that is equal to $(4+N_u)$.

\begin{figure}[t!]
\centerline{\includegraphics [scale = 0.5] {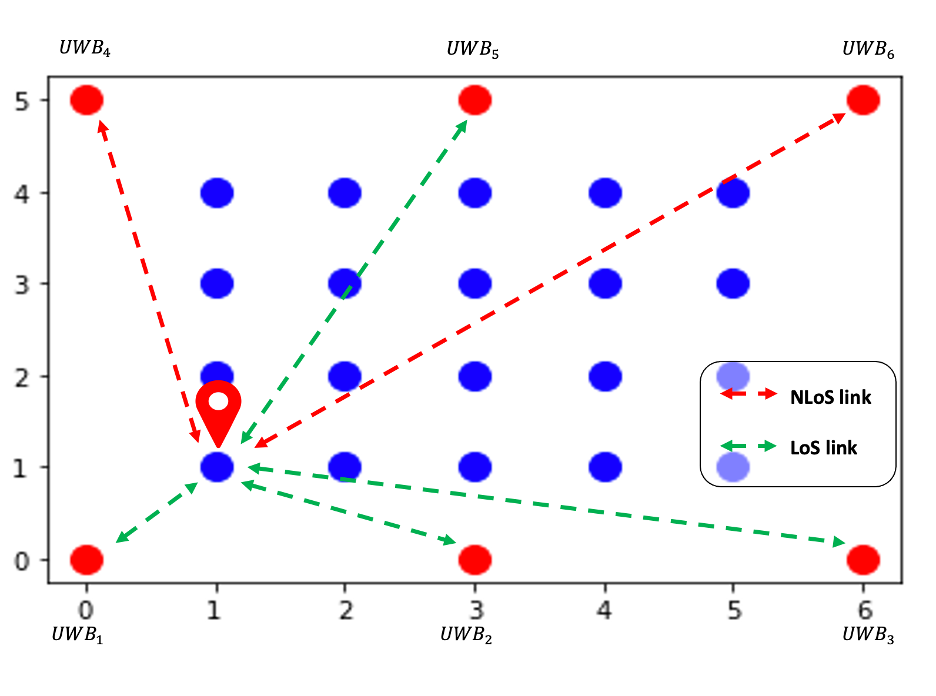}}
\caption{\footnotesize A typical sub-area in an experimental indoor environment consisting of $6$ UWB beacons for UWB node selection.}
\label{Fig5}
\end{figure}
\begin{table}[t]
\centering
\caption{\footnotesize List of Parameters.}\label{tab2}
\begin{tabular}{ |c|c|c||c|c|c|}
\hline
\textbf{Notation} & \textbf{Unit} & \textbf{Value} & \textbf{Notation} & \textbf{Unit} &\textbf{Value} \\ \hline
$L_{SP}$ & symbols& 1024  & $P_{SYN}$ &mW&  30.6\\
$L_{PHR}$ & symbols& 16  & $P_{ADC}$ &mW& 2.2\\
$N_p $ & -& $\{1, 3, 5, \ldots, 15\}$ &$P_{GEN}$ &mW& 2.8\\
$E_p $ &pJ/pulse& 4.5 &$P_{LNA}$ &mW& 9.4\\
$R_{base}$& Mbps& 1&$P_{EST}$ &mW& 10.08 \\
$P_{COR}$ &mW& 10.08  & $P_{VGA}$& mW& 22\\ \hline
\end{tabular}
\end{table}
%
\vspace{-.1in}
\section{Evaluations and Simulation Results}\label{Sec:4}
\begin{figure}[t!]
\centering
\mbox{\subfigure[$N_u=4$]{\includegraphics[scale = .2]{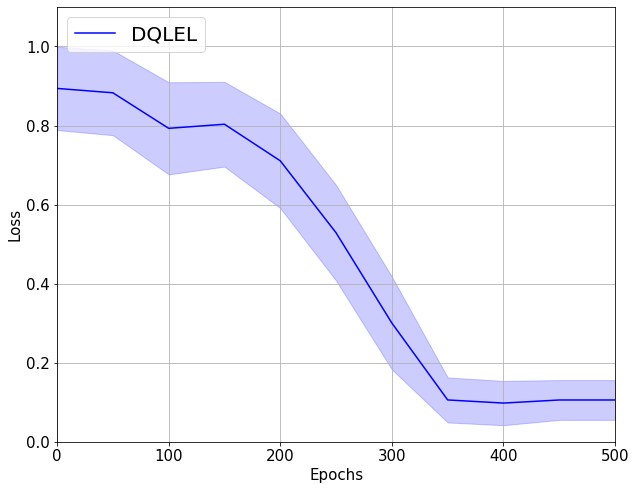}}
\subfigure[$N_u=6$]{\includegraphics[scale = .2]{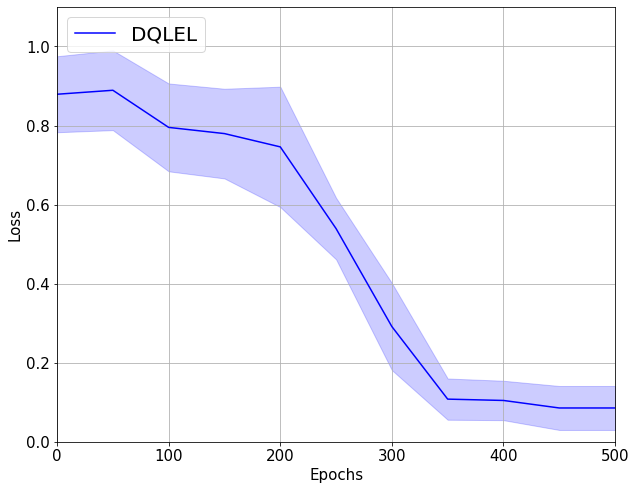}}}
\caption{\footnotesize The convergence of the proposed DQLEL framework for $(a)$ $N_u=4$, and $(b)$ $N_u=6$.}\label{Fig:thirteen}
\end{figure}
\begin{figure}[t!]
\centering
\mbox{\subfigure[$N_u=4$]{\includegraphics[scale = .2]{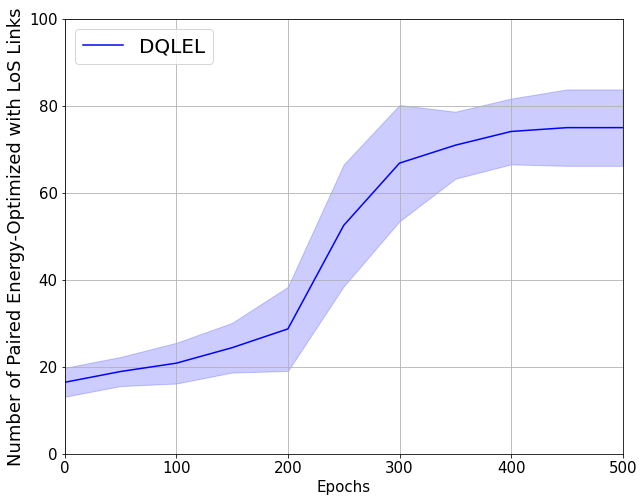}}
\subfigure[$N_u=6$]{\includegraphics[scale = .2]{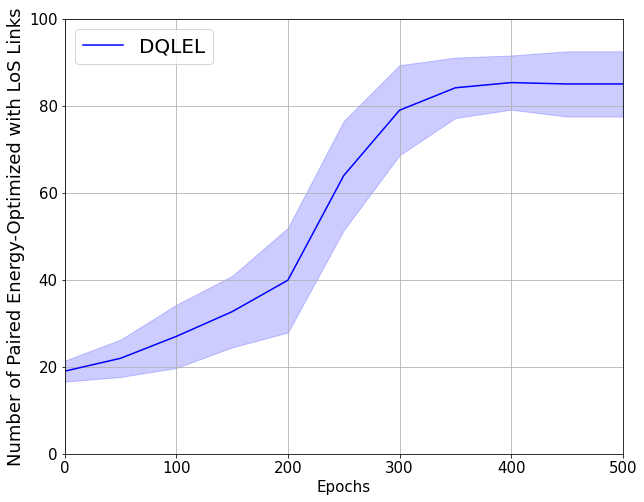}}}
\caption{\footnotesize Number of paired energy-optimized with LoS Links $(a)$ $N_u=4$, and $(b)$ $N_u=6$.}\label{fig:six}
\end{figure}
To evaluate the effectiveness of the proposed DQLEL scheme, we consider an experimental indoor testbed with the size of $(60 \times 50)$ $m^2$. Indoor environments, such as hotels and office buildings, can be divided into several non-overlapping sub-areas~\cite{Wang:2016, Li:2020}, where the mobile user is localized in each sub-area through all that sub-area's UWB beacons. Due to the limited transmission area of UWB beacons, after transmitting an UWB signal by the user at each location, the transmitted signal can be received by $N_u$ number of UWB beacons. Fig.~\ref{Fig5} illustrates a typical sub-area consisting of six UWB beacons, located at a $(6 \times 5)$ $m^2$ rectangular indoor environment. Following Reference~\cite{Li:2020}, the sub-area is divided into ($5 \times4$) square zones with dimension of ($1 \times 1$) $m^2$. The mobile user moves across the environment in $8$ directions according to Eqs.~\eqref{e19} and~\eqref{e20} based on the random walk model~\cite{Li:2020}. At each time slot, the mobile user is located at the center of each zone~\cite{Li:2020}, shown by a blue point in Fig.~\ref{Fig5}. The channel condition of the received signal depends on the LoS/NLoS of the signal. To consider effects of obstacles  on the received signal, the channel condition of $UWB_i$, for ($1\leq i \leq N_u$), at each location is determined randomly. Two out of $N_u$ beacons will be selected for localization. Table~\ref{tab2} illustrates the list of other parameters~\cite{Wang:2010} used for running the experiments.
Taking the above considerations into account, first, in Sub-section~\ref{subsec:Eff}, we evaluate the effectiveness of the proposed DQLEL framework. Then in Sub-section~\ref{subsec:Comp}, we compare the performance of the proposed DQLEL framework with conventional schemes from the aspect of location accuracy, the battery-life of UWB beacons, and the channel condition of candidate beacons.

\vspace{-0.1in}
\subsection{Effectiveness of the DQLEL Framework} \label{subsec:Eff}
\begin{figure}[t!]
\centering
\mbox{\subfigure[$N_u=4$]{\includegraphics[scale = .2]{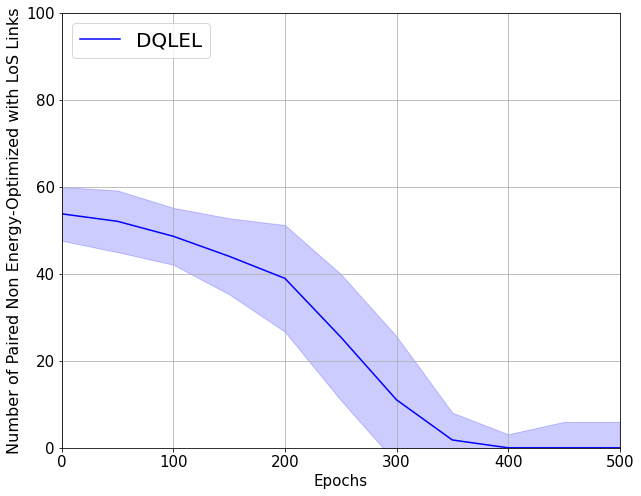}}
\subfigure[$N_u=6$]{\includegraphics[scale = .2]{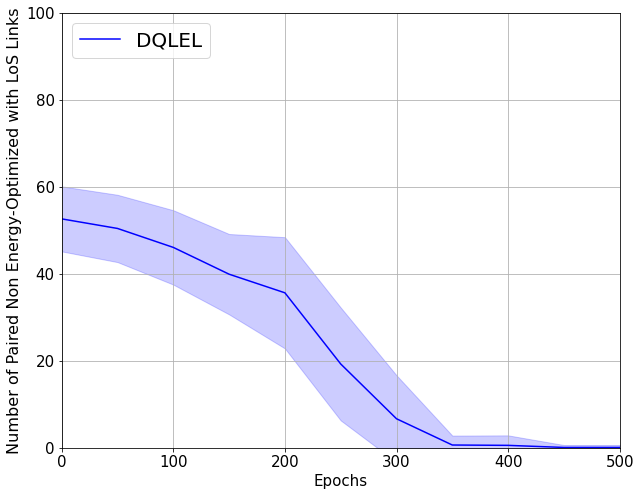}}}
\vspace{-.15in} \caption{\footnotesize Number of paired non energy-optimized with LoS links for $(a)$ $N_u=4$, and $(b)$ $N_u=6$.}\label{fig:seven}
\end{figure}

\begin{figure}[t!]
\centering
\mbox{\subfigure[$N_u=4$]{\includegraphics[scale = .2]{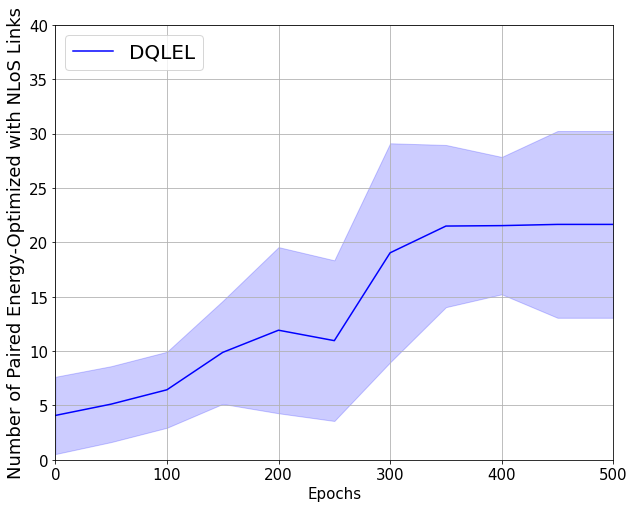}}
\subfigure[$N_u=6$]{\includegraphics[scale = .2]{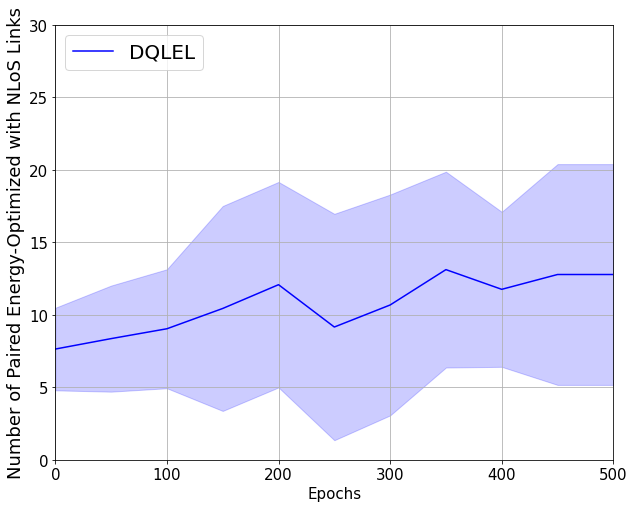}}}
\vspace{-.1in} \caption{\footnotesize Number of paired energy-optimized with NLoS links for $(a)$ $N_u=4$, and $(b)$ $N_u=6$.}\label{fig:eight}
\end{figure}

\begin{figure}[t!]
\centering
\mbox{\subfigure[$N_u=4$]{\includegraphics[scale = .2]{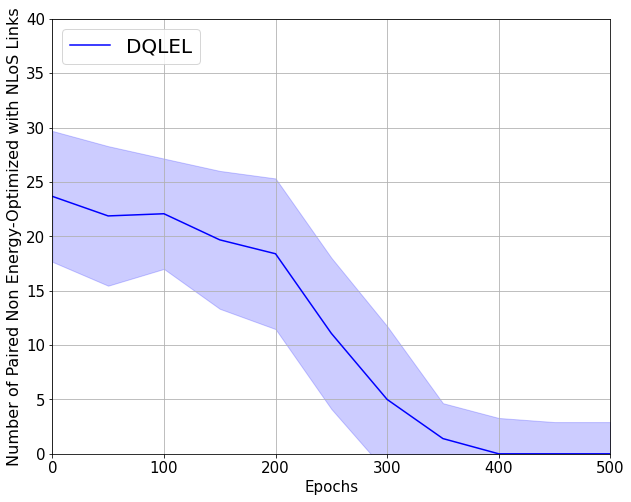}}
\subfigure[$N_u=6$]{\includegraphics[scale = .2]{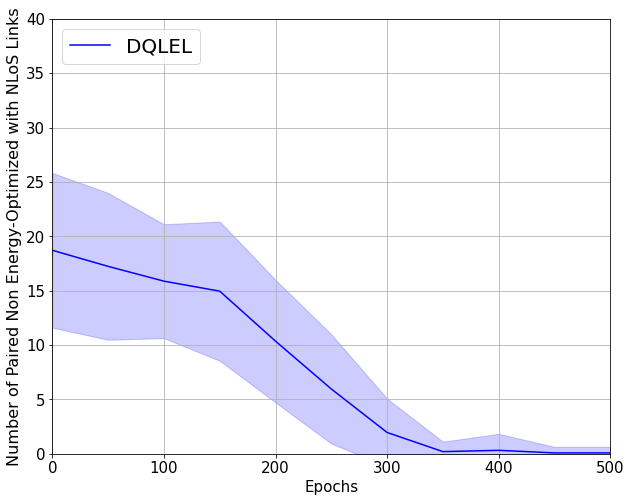}}}
\vspace{-.1in} \caption{\footnotesize Number of paired non energy-optimized with NLoS links for $(a)$ $N_u=4$, and $(b)$ $N_u=6$.}\label{fig:nine}
\end{figure}

Convergence  of  the  proposed DQLEL framework is evaluated in Fig.~\ref{Fig:thirteen}. According to Eq.~\eqref{e24}, the learning process is performed by minimizing the loss function, which is the mean-squared error of the target optimal Q-function with the minibatch updates. Fig.~\ref{Fig:thirteen} illustrates that the proposed DQLEL framework converges after $350$ epochs.  
Moreover, Figs.~\ref{fig:six}-\ref{fig:nine} illustrate the number of connections of each type in different epochs. We also investigate the effect of the number of UWB beacons in each sub-area in Figs.~\ref{fig:six}-\ref{fig:nine}.

Fig.~\ref{fig:six} illustrates the number of paired energy-optimized with LoS links in the proposed DQLEL framework for $N_u=4$ and $N_u=6$. It is worth mentioning that given the limited transmission range of the UWB beacons, it is common to have a small number of beacons in each sub-area, typically, $4$ beacons are considered. To evaluate the effects the number of beacons on the DQLEL framework, we consider the common scenario with $4$ beacons together with a second scenario with $6$ UWB beacons. The environment for both scenarios is considered to be similar, therefore, the number of NLoS links in both cases would be the same, while the number of possible actions in $N_u=4$ and $N_u=6$ are $6$ and $15$, respectively. Under the assumption that in each location, at least one link is NLoS, then the probability of establishing LoS connections for $N_u=4$ and $N_u=6$ are $50\%$ and $66\%$, respectively. Intuitively speaking, this means that when the action space is small (e.g., $6$ actions when there are $4$ beacons compared to $15$ actions when we have two extra beacons), it is less likely to have paired LoS links. According to the results in Fig.~\ref{fig:six}, it can be observed that the number of energy-optimized with LoS connections increases as the number of epochs grows and converges after about $330$ epochs to $75\%$ and $85\%$ for $N_u=4$ and $N_u=6$, respectively.

Fig.~\ref{fig:seven} illustrates the number of paired LoS links that make the deviation of remaining battery life of UWB beacons become greater than $\mathcal{MD}_{th}$. As it can be seen from Figs.~\ref{fig:six} and~\ref{fig:seven}, most LoS links at the earlier epochs are non energy-optimized. By increasing the epochs, however, the number of non energy-optimized LoS links decreases resulting in more energy-optimized LoS connections. According to the results in Fig.~\ref{fig:seven}, the number of non energy-optimized links converges to zero after about $350$ epochs in both $N_u=4$ and $N_u=6$.

\begin{figure}[t!]
\centering
\mbox{\subfigure[$N_u=4$]{\includegraphics[scale = .2]{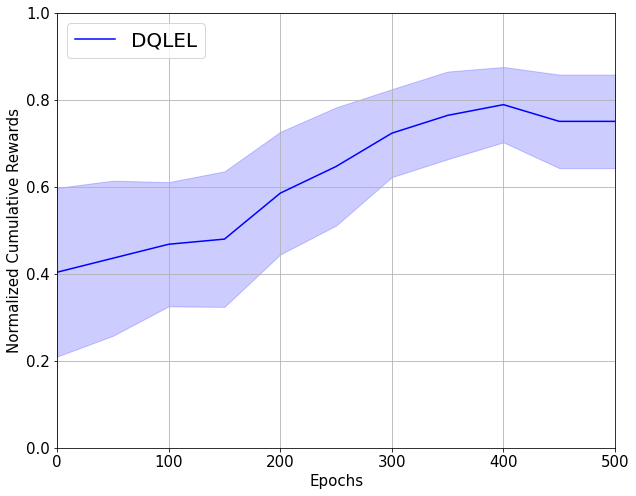}}
\subfigure[$N_u=6$]{\includegraphics[scale = .2]{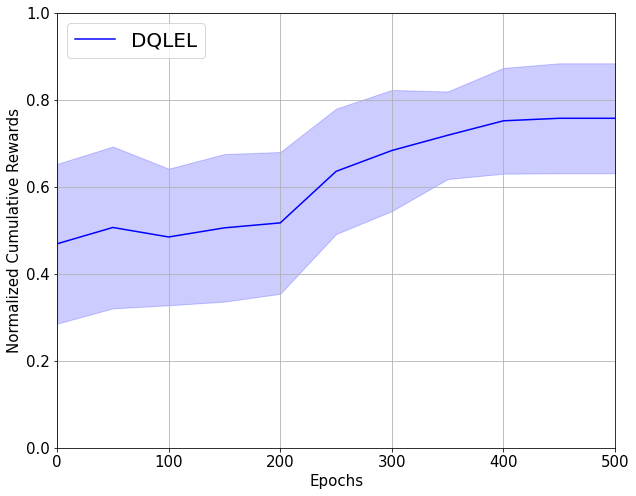}}}
\caption{\footnotesize The variation of normalized cumulative rewards versus different epochs for $(a)$ $N_u=4$, and $(b)$ $N_u=6$.}\label{Fig:ten}
\end{figure}

The number of energy-optimized and non energy-optimized NLoS connections are depicted in Figs.~\ref{fig:eight} and~\ref{fig:nine}, respectively. By a similar argument, non energy-optimized NLoS links in Fig.~\ref{fig:nine} experience a remarkable reduction, eventually converging to zero after $350$ epochs, which results in a slight increase in the energy-optimized NLoS connections. By considering the fact that the proposed DQLEL framework needs to maintain a trade-off between two objectives, i.e., the link condition and energy consumption of UWB beacons, the small growth in number of energy-optimized NLoS connections in Fig.~\ref{fig:eight} is acceptable. Fig.~\ref{Fig:ten} illustrates the normalized cumulative rewards of the agent in each epoch. According to the reward definition in Eq.~\eqref{e23}, energy-optimized LoS connections result in a considerable increase in the cumulative rewards. According to the results in Fig.~\ref{Fig:ten}, increasing the number of epochs increases the cumulative rewards, showing that the model is well-trained.

\begin{figure}[t!]
\centering
\mbox{\subfigure[$N_u=4$]{\includegraphics[scale = .2]{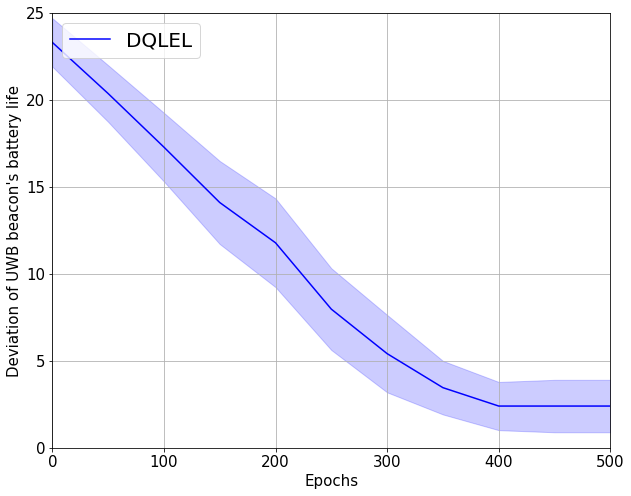}}
\subfigure[$N_u=6$]{\includegraphics[scale = .2]{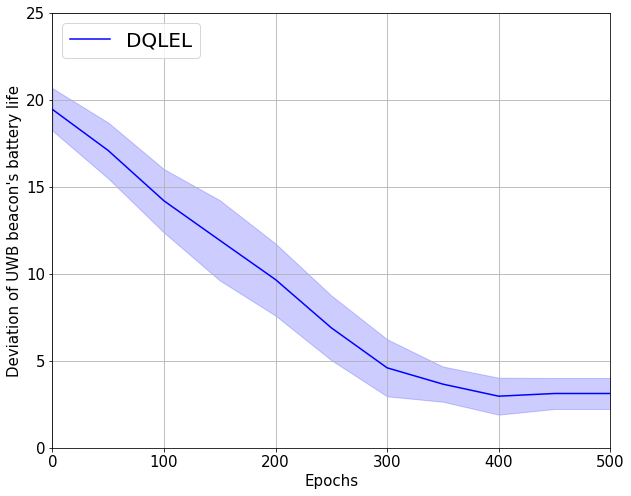}}}
\caption{\footnotesize Deviation of UWB beacons battery life versus different epochs for $(a)$ $N_u=4$, and $(b)$ $N_u=6$.}\label{Fig:eleven}
\vspace{-.1in}
\end{figure}
\begin{figure}[t!]
\centering
\mbox{\subfigure[$N_u=4$]{\includegraphics[scale = .2]{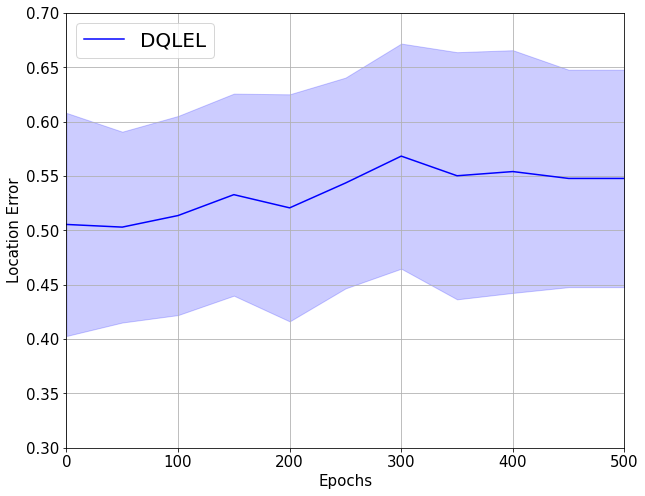}}
\subfigure[$N_u=6$]{\includegraphics[scale = .2]{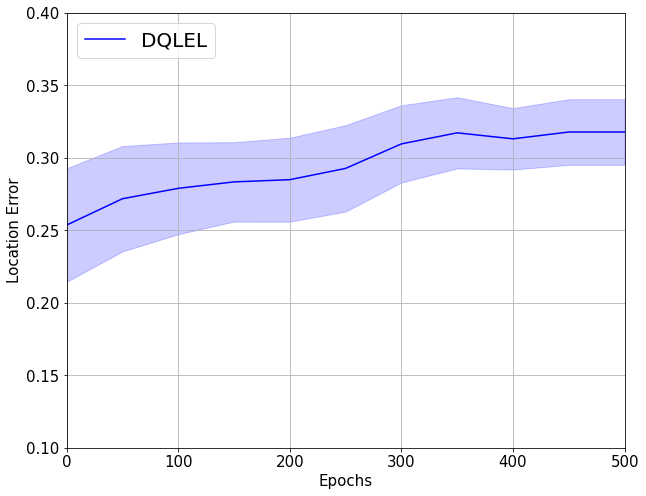}}}
\vspace{-.1in} \caption{\footnotesize Location error (m) versus different epochs for $(a)$ $N_u=4$, and $(b)$ $N_u=6$.}\label{Fig:twelve}
\end{figure}

\vspace{-.15in}
Fig.~\ref{Fig:eleven} evaluates the deviation of the UWB beacons' battery life, obtained by Eq.~\eqref{e22}, versus different epochs for $(a)$ $N_u=4$, and $(b)$ $N_u=6$. Note that, larger value of batteries' deviation indicates that certain UWB beacons are involved in localization more than others. In this model, we assume that each epoch is terminated if the current time reaches a pre-determined time value, or the battery of at least one UWB beacon is completely drained before the time threshold. Without considering the UWB beacon's energy consumption, the agent identifies a pair of UWB beacons with LoS links, draining the battery of those corresponding beacons by being repeatedly selected. The main goal of the proposed DQLEL framework is to select UWB beacons with LoS links while minimizing the deviation of UWB beacons' battery lives. As it can be seen from Fig.~\ref{Fig:eleven}, the deviation of UWB beacons' battery lives decreases as the number of epochs grows and converges after about $350$ epochs to $2.4$ and $3.8$ for $N_u=4$ and $N_u=6$, respectively. In addition, Fig.~\ref{Fig:twelve} illustrates the location error versus different epochs, which is obtained by Eq.~\eqref{e18}. As it can be seen from Fig.~\ref{Fig:twelve}, the location error slightly increases by increasing the number of epochs, which is negligible. The main reason behind this is that there is a trade-off between the number of LoS links (accuracy) and the deviation of UWB beacons battery life. This completes the evaluation of the DQLEL framework.

\begin{figure*}[t!]
\centering
\mbox{\subfigure[]{\includegraphics[scale = .19]{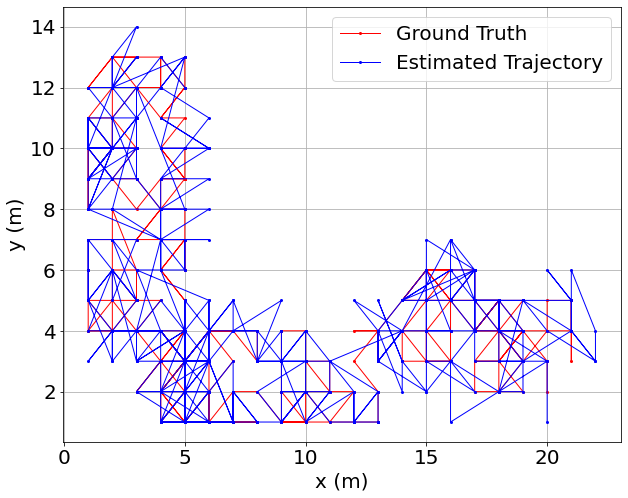}}
\subfigure[]{\includegraphics[scale = .19]{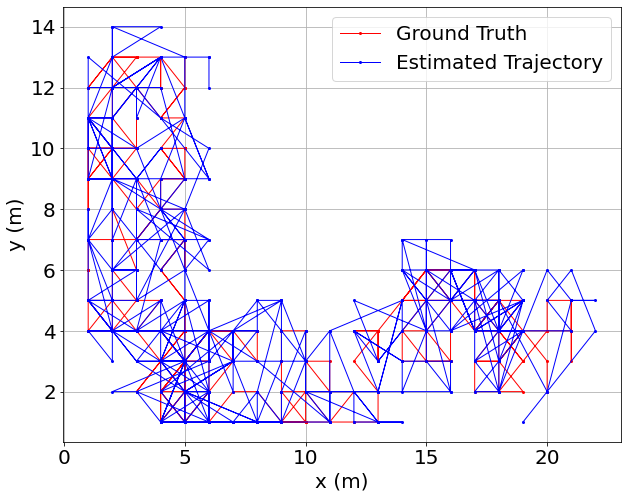}}
\subfigure[]{\includegraphics[scale = .19]{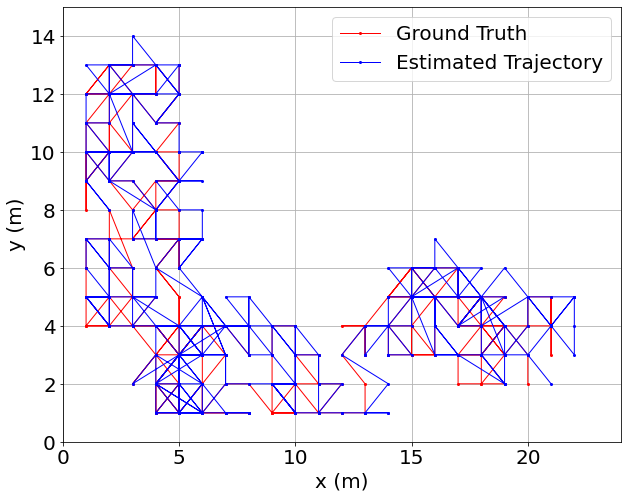}}
\subfigure[]{\includegraphics[scale = .19]{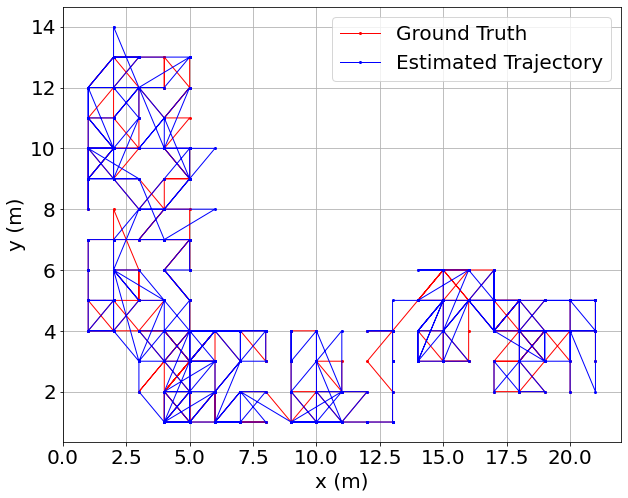}}}
\vspace{-.15in}
\caption{\footnotesize Ground truth and estimated random trajectories by using (a)  NN-NS, (b) RNS, (c) NE-DRL, and (d) DQLEL frameworks.}\label{Fig:12}
\vspace{-.1in}
\end{figure*}

\begin{figure*}[t!]
\centering
\mbox{\subfigure[]{\includegraphics[scale = .2]{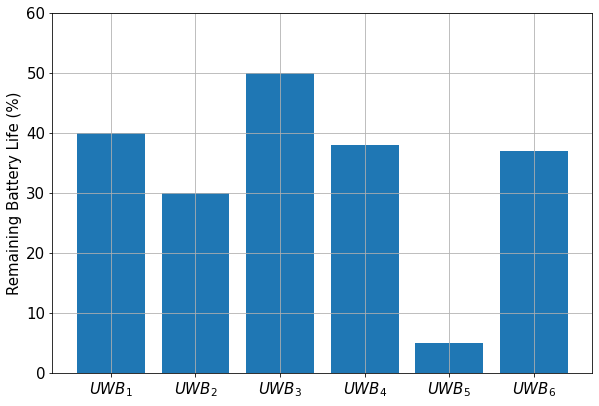}}
\subfigure[]{\includegraphics[scale = .2]{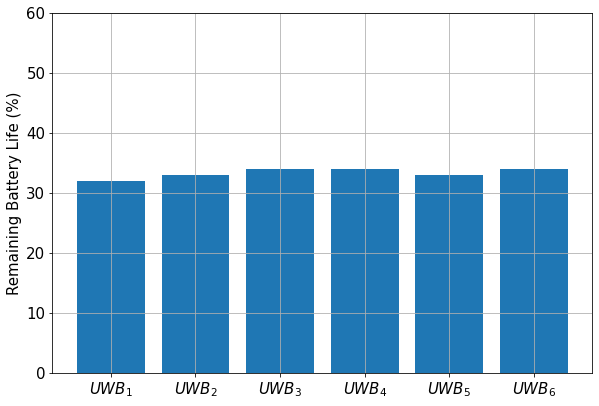}}
\subfigure[]{\includegraphics[scale = .2]{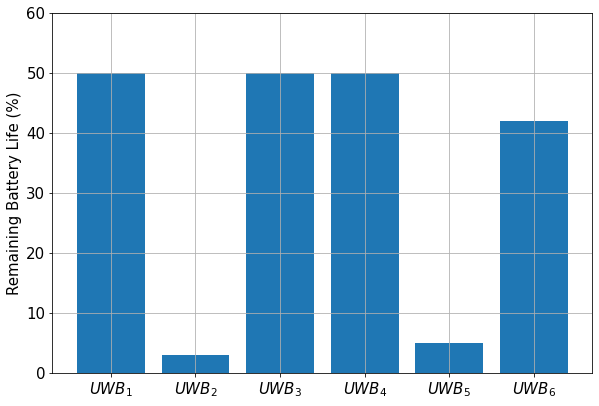}}
\subfigure[]{\includegraphics[scale = .2]{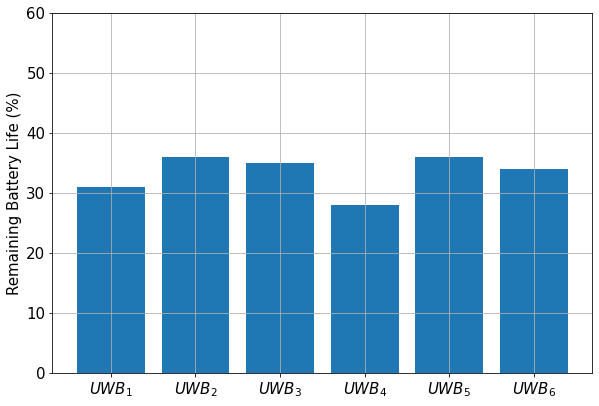}}}
\vspace{-.15in}
\caption{\footnotesize Percentage of remaining battery life of beacons in one sub-area after $50$ iterations by using (a)~NN-NS, (b) RNS, (c) NE-DRL, and (d) DQLEL.}\label{Fig:13}
\vspace{-.1in}
\end{figure*}
\vspace{-0.1in}
\subsection{Performance Comparisons} \label{subsec:Comp}
To the best of our knowledge, there is no RL-based UWB node selection framework that  studied both LoS/NLoS and energy perspectives, for comparison purposes. Therefore, we introduce three baseline models for comparisons:
\begin{itemize}
\item \textit{\textbf{Non Energy-Optimized DRL (NE-DRL) LoS/NLoS UWB Node Selection:}} Similar to the proposed DQLEL framework, i.e., all the parameters are the same, with the difference that in this baseline model, we just consider the link condition as the reward function. For an action that both UWB links are LoS connections, the reward $r_t$ is set equal to $1$, otherwise, it equals $-1$.
\item \textit{\textbf{Random UWB Node Selection (RNS):}} In this framework, a pair of UWB beacons are randomly selected  for localization without considering the remaining battery life of UWB beacons and channel conditions.
\item \textit{\textbf{Nearest Neighbor UWB Node Selection (NN-NS):}} Similar to the previous one, without considering the remaining battery life of UWB beacons and channel conditions, two nearest UWB beacons are selected for localization.
\end{itemize}
For comparison purposes, we consider a $(24 \times 15)$ $m^2$ rectangular indoor environment. Fig.~\ref{Fig:12} illustrates the estimated random trajectory of the mobile user by using NN-NS, RNS, NE-DRL, and DQLEL frameworks. Since the mobile user in the NE-DRL and DQLEL frameworks is trained to be localized by LoS beacons, the estimated trajectory is almost the same as the ground truth (see Fig.~\ref{Fig:12}). Similarly, Fig.~\ref{Fig:13} compares the remaining battery life of UWB beacons after $50$ iterations. According to the results in Fig.~\ref{Fig:13}, the battery of certain UWB beacons are completely drained without applying the DQLEL framework, while the remaining battery of UWB beacons are almost the same in the DQLEL framework, leading to a remarkable increase in the life time of the~infrastructure.

\begin{figure*}[t!]
\centering
\mbox{\subfigure[]{\includegraphics[scale = .22]{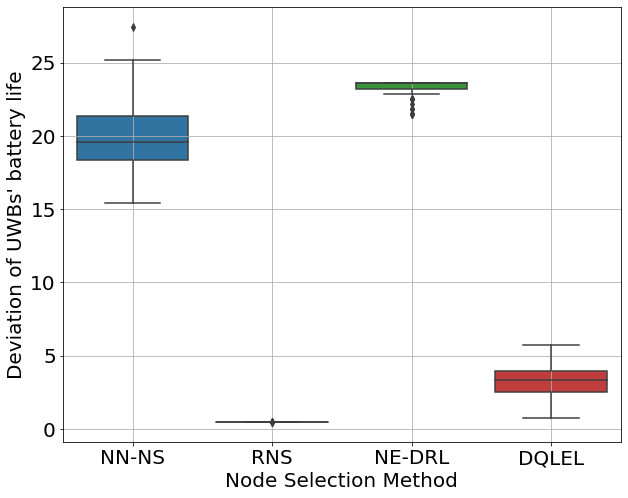}}
\subfigure[]{\includegraphics[scale = .22]{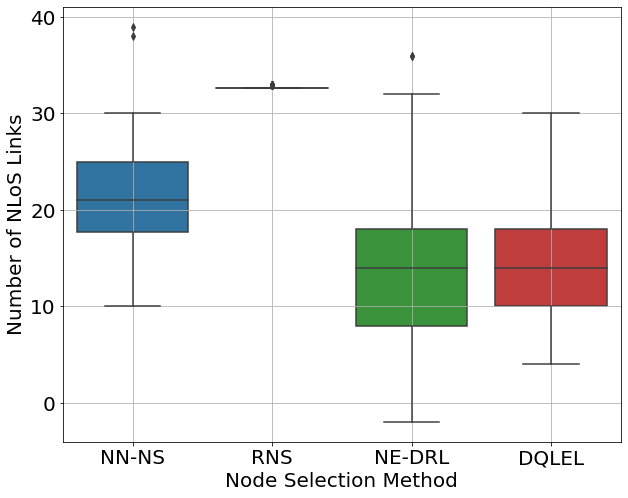}}
\subfigure[]{\includegraphics[scale = .22]{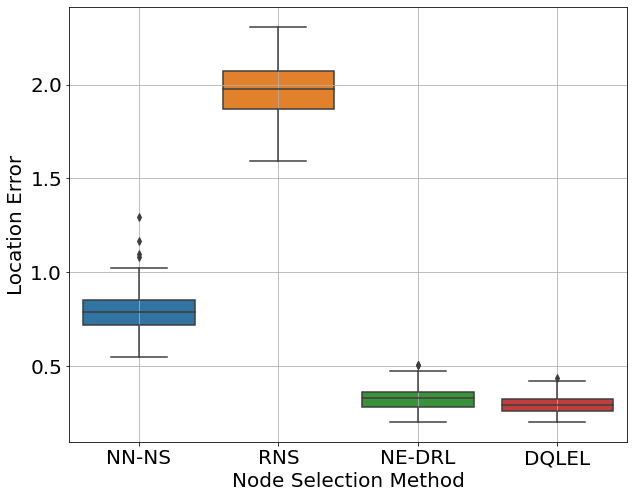}}}
\vspace{-.15in}
\caption{\footnotesize (a) A comparison of different node selection methods based on (a) Deviation of UWBs' battery life; (b) Number of NLoS links; (c) Location error.}
\label{Fig14}\label{Fig15}\label{Fig16}
\end{figure*}

\begin{figure}[t!]
\centerline{\includegraphics [scale = 0.23] {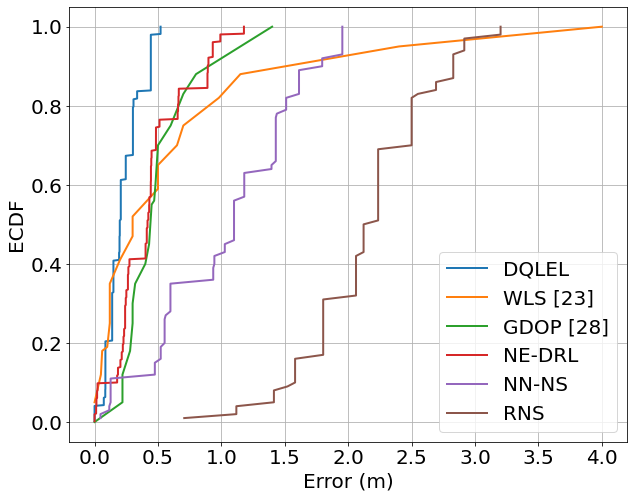}}
\vspace{-.1in}
\caption{\footnotesize A comparison with state-of-the-arts based on the location error ECDF.}
\label{Fig17}
\end{figure}
Fig.~\ref{Fig14}(a) compares the performance of the proposed DQLEL framework with NE-DRL, RNS, and NN-NS schemes in terms of the deviation of UWBs' battery life. Due to the random movement of the mobile user and the random UWB node selection approach of the RNS framework, the remaining battery life of the UWB beacons are almost the same. Therefore, the mean deviation of UWBs' battery life is almost zero. In other words, the RNS approach performs reasonably well in terms of energy distribution, however, fails to accurately localize the target.  In contrary to the RNS method, the mean deviation of UWBs' battery life of the NN-NS and NE-DRL is considerably high. The main reason behind this high mean deviation of UWBs' battery life  is that the NN-NS and NE-DRL frameworks select the nearest beacons and the beacons with LoS links for localization, respectively, without taking into account energy considerations. As a result, the battery of certain UWB beacons are completely drained leading to a high deviation of UWBs' battery life. The proposed DQLEL framework, however, provides a low battery life deviation, while enhancing the location accuracy by choosing UWB beacons with LoS~links.

Fig.~\ref{Fig15}(b) illustrates the number of NLoS links established by different node selection methods. Note that, using less number of NLoS links provides the higher location accuracy. As it can be seen from Fig.~\ref{Fig15}(b), the proposed DQLEL framework and NE-DRL experience lower number of NLoS links. RNS method, however,  results in the highest number of NLoS connections since there is no link condition criteria. By considering the fact that the nearest UWB beacons provide LoS links with a higher probability, the possibility of selecting UWB beacons with NLoS links in the NN-NS scheme is lower than the RNS~method. Fig.~\ref{Fig16}(c) compares the performance of the proposed DQLEL framework with NE-DRL, RNS, and NN-NS schemes from the aspect of the location error. By a similar argument, which is used for the number of NLoS links, localization by applying RNS method leads to the highest location error. The proposed DQLEL and NE-DRL frameworks have the lowest position error since they both select LoS connections. Finally, to demonstrate the superiority of the proposed DQLEL framework in comparison to its state-of-the-art counterparts, we calculate the Empirical Cumulative Distribution Function (ECDF) of the WLS~\cite{Albaidhani:2019}, GDOP~\cite{Albaidhani:2020}, NE-DRL (which itself is a simplified version of the DQLEL framework), RNS, and NN-NS  anchor selection schemes. According to the results shown in Fig.~\ref{Fig17}, the location error of the DQLEL framework is lower than other methods. 


\vspace{-.1in}
\section{Conclusion}\label{Sec:5}
Within the context of UWB-based indoor localization, the paper proposed the Deep Q-Learning Energy-optimized LoS/NLoS  (DQLEL) UWB node selection framework to maintain a balance between the remaining battery life of UWB beacons and localization accuracy.
The mobile user running the DQLEL framework is autonomously trained to determine the optimal pair of UWB beacons to be localized based on the 2-D TDoA framework. The effectiveness of the proposed DQLEL framework is evaluated in terms of the link condition, the deviation of the remaining battery life of UWB beacons, location error, and cumulative rewards. Simulation results showed that the proposed DQLEL framework improved the location accuracy by considering the link condition as the selection criteria, increased the number of selected UWB beacons with LoS links, and provided a balance between the remaining battery life of all UWB beacons. Based on the simulation results, the proposed DQLEL framework illustrates significant performance improvements in comparison to its counterparts across the aforementioned aspects. In this paper, as the first step towards development of a fully autonomous agent for UWB-based indoor localization, we considered one active user in each time slot. With the emphasis on the multiple access technologies, our future research involves the deployment of a DRL-based localization framework capable of localizing multiple users in each time~slot.

\bibliographystyle{IEEEtran}
\bibliography{keylatex}

\end{document}